\documentclass[10pt]{revtex4}
\usepackage{graphicx}
\usepackage{amsthm} 

\begin{document}

\title{An Information--Theoretic Equality Implying the Jarzynski Relation}
\author{Vlatko Vedral}
\affiliation{Clarendon Laboratory, University of Oxford, Parks Road, Oxford OX1 3PU, United Kingdom\\Centre for Quantum Technologies, National University of Singapore, 3 Science Drive 2, Singapore 117543\\
Department of Physics, National University of Singapore, 2 Science Drive 3, Singapore 117542}

\begin{abstract}
We derive a general information-theoretic equality for a system undergoing two projective measurements separated by a general temporal evolution. The equality implies the non-negativity of the mutual information between the measurement outcomes of the earlier and later projective measurements. We show that it also contains the Jarzynski relation between the average exponential of the thermodynamical work and the exponential of the difference between the initial and final free energy. Our result elucidates the information-theoretic underpinning of thermodynamics and explains why the Jarzynski relation holds identically both quantumly as well as classically.    
\end{abstract}

\maketitle

The sole purpose of this letter is to derive a simple information-theoretic equality for a general quantum mechanical process that involves a state preparation (by a projective measurement), followed by a general quantum evolution of the system, and finally concluded by executing another projective measurement. This scenario is completely general and effectively captures any question that can be posed within the formalism of quantum physics. We introduce no other constraints for the time being. To make our analysis quantitative, let us start in the state $\rho$, and let us label the basis of the first measurement $\{P_n\}$. The ensuing evolution will be given by the most general completely positive, trace preserving map, $\sum_i \Lambda_i (\cdot) \Lambda^{\dagger}_i$. The basis of the final measurement will be $\{Q_m\}$. The main quantity, from which all other physically relevant quantities can be derived, will be the joint probability $p(n,m)$ that the system is initially projected onto the state $P_n$ and then also projected onto the state $Q_m$ in the final measurement. It is given by the Born rule
\begin{equation}
p(n,m) = tr \{Q_m \sum_i \Lambda_i P_n \rho P_n \Lambda^{\dagger}_i Q_m\} = tr Q_m (\sum_i \Lambda_i P_n \Lambda^{\dagger}_i)\times tr (P_n \rho)\; .
\end{equation}   
This formula can be understood as a simple expression of the usual probabilistic chain rule $p(n,m) = p(m|n) p(n)$, where
\begin{equation}
p(n) = tr (P_n \rho)\; 
\end{equation} 
is the probability for the $n$th outcome in the first measurement, while
\begin{equation}
p(m|n) = tr Q_m (\sum_i \Lambda_i P_n \Lambda^{\dagger}_i)
\end{equation}
is the conditional probability that the second measurement will yield the outcome $m$ if the first measurement was $n$. The key quantity in our equality will be the mutual information \cite{Shannon} between $P_n$ and $Q_m$, denoted by $I_{nm}$, which we now proceed to define as
\begin{equation}
I_{nm} = -\ln p(m) + \ln p(m|n) \; ,
\end{equation}
where $p(m) = \sum_{n} p(n,m) = tr Q_m \Lambda_i \rho \Lambda^{\dagger}_i$. The interpretation of this quantity is simple. In the single realisation of our experiment, when the measurement outcomes are first $P_n$ and then $Q_m$, the mutual information tells us about the difference between the entropy of the $m$th outcome without the knowledge of $n$ (given by $-\ln p(n)$) and the $m$th outcome when $n$ is known (given by $-\ln p(m|n)$). This quantity measures correlations between two measurements and is symmetric under the exchange of the two measurements (which is why temporal labeling is actually superfluous).  

The information-theoretic equality presenting our main result can now be stated as:
\begin{equation}   
\langle e^{-I_{nm}} \rangle := \sum_{nm} p(n,m) e^{-I_{nm}} = 1 \; .
\end{equation}
Expressed in words, the average of the exponential of the single trial mutual information between the preparation and measurement on a system free to evolve in a general way between the two, always equals a unity. 

The proof of the above equality is simple. By substituting the expression of $I_{nm}$ into the above formula we obtain
\begin{equation}   
\sum_{nm} p(n,m) e^{-I_{nm}} = \sum_{nm} p(n,m) e^{\ln p(m) - \ln p(m|n)} = \sum_{nm} p(n,m) \frac{p(m)}{p(m|n)} = \sum_{nm} p(n)\times p(m) = \sum_n p(n)\times \sum_m p(m) = 1 \; .
\end{equation}
Therefore, following this chain of equalities backward, we can conclude that our information-theoretic equality is a simple expression of the conservation of probability. 

We conclude by presenting two consequences of the above equality. The first is that the equality implies the non-negativity of the total mutual information $\sum_{nm} p(n,m) I_{n,m} \geq 0$ (and is therefore a stronger statement), which just follows from the convexity of the exponential function. This may seem like a trivial statement, however, it contains insightful results. For instance, the state $\rho$ can be thought of as being bi-partite, and the first and last measurements could contain projections on both subsystems. Our equality could then be applied to the trade-off between the mutual informations between subsystems and the sum of the entropy changes of the subsystems (see the Maxwell's demon discussion in \cite{Penrose}).   

The second implication of our equality is the Jarzynski relation. For this we need to assume that the initial state $\rho$ is a Gibbs state of the form $\rho = \exp \{-\beta H\} / Z$ where $Z = tr \exp \{-\beta H\}$ is the partition function and $H=\sum_n E_n P_n$ (as usual $\beta = 1/kT$, $k$ being Boltzmann's constant and $T$ the temperature). Secondly, the basis of the first measurement is the same as that of $H$ (and therefore of $\rho$). Thirdly, the evolution is generated by a Hamiltonian $H'=\sum_m E'_m Q_m$ and the second measurement is performed in the basis of $H'$. Then it can be shown that the mutual information can be written as $I_{nm} = \beta (E_n-E'_m)- \beta (F' - F)$ where $F =  -kT \ln Z$ and $F' =  -kT \ln Z'$ are the initial and the final free energies and $Z' = tr \exp \{-\beta H'\}$ is the final partition function (here we assume the same initial and final temperatures for simplicity, but without any real loss of generality). Therefore, our information-theoretic equality can now be written as 
\begin{equation}   
\langle e^{-I_{nm}} \rangle = \langle e^{\beta(E_n-E'_m)-\beta (F'-F)} \rangle = 1 \; .
\end{equation}
which is, in fact, the quantum version of the Jarzynski equality \cite{Jarzynski,Tasaki}. Interestingly, this allows us to interpret the quantity $ \beta (E_n-E'_m)- \beta (F' - F)$ (energy change minus free energy change between two measurements) as the mutual information, and therefore correlation, between the past and future measurements (or between the initial preparation and the final measurement).  

In summary, we have presented a very general equality involving the exponential of the mutual information whose validity rests purely on the simple properties of classical probabilities. We have demonstrated its power by showing that it implies a (restricted) version of the quantum Jarzynski equality (the fact that even though we considered a general quantum setting all that mattered were classical probabilities explains why the Jarzynski equality is true both in quantum as well as in classical physics). It will not be difficult to demonstrate that more general scenarios, including errors and feedback (i.e. cases where $\Lambda$s become dependent on $n$), can also be shown to follow from the presented information-theoretic equality, a task we leave to subsequent investigations.  

\textit{Acknowledgments}: The author acknowledges financial support from the Templeton foundation, Leverhulme Trust (UK), as well as the National Research Foundation and Ministry of Education, in Singapore. The author is a fellow of Wolfson College Oxford.

\end{document}